\documentstyle[12pt]{article}
\begin{document}
\parindent 1,2 \parindent
\pagestyle{plain}
\baselineskip 1.3 \baselineskip


\newcommand{\be}{\begin{equation}}
\newcommand{\ee}{\end{equation}}
\newcommand{\bea}{\begin{eqnarray}}
\newcommand{\eea}{\end{eqnarray}}
\newcommand{\bec}{\begin{center}}
\newcommand{\eec}{\end{center}}
\newcommand{\del}{\tilde{\nabla}}
\newcommand{\boxy}{\widetilde{\Box}}
\newcommand{\tg}{\tilde{g}}
\newcommand{\tr}{{\rm tr}}
\newcommand{\Tr}{{\rm Tr} }
\newcommand{\no}{\nonumber}

\newcommand{\bph}{\bar{\phi}}
\newcommand{\hph}{\hat{\phi}}
\newcommand{\idg}{\int_\Sigma d^3 x \sqrt{g}}
\newcommand{\idgmx}{\int_{\cal M} d^4 x \sqrt{g}}
\newcommand{\idgmy}{\int_{\cal M} d^4 y \sqrt{g}}
\newcommand{\V}{\tilde{V}}
\newcommand{\pole}{ - \frac{1}{\epsilon} + \frac{1}{2} \gamma
		  + \ln (\frac{ \beta^2 M^2}{ 4 \pi} ) }
\newcommand{\polea}{ - \frac{1}{\epsilon} + \frac{1}{2} \gamma
		  + \ln (\frac{  M^2}{ 4 \pi {\tilde{\mu}}^2} )}
\newcommand{\Mp}{ M^2 - \mu^2 + \frac{ \lambda}{6} \bph^2 }
\newcommand{\Mm}{ M^2 - \mu^2 - \frac{ \lambda}{6} \bph^2 }
\newcommand{\Kit}{  \frac{1}{6} ( \bph_1^2 - \bph_2^2) \Box
		    ( \bph_1^2 - \bph_2^2 ) + \frac{2}{3}
		    \bph_1 \bph_2 \Box ( \bph_1 \bph_2) }
\newcommand{\half}{\frac{1}{2}}
\newcommand{\Nhalf}{\frac{N}{2}}

\begin{flushright}
KAIST-CHEP-93/M4
\end{flushright}
\begin{center}
{\bf{ \Large  Bose-Einstein condensation for a self-interacting
theory in  curved spacetime }}
\end{center}
\vspace{1cm}
\begin{center}
Min-Ho Lee, Hyeong-Chan Kim, and Jae Kwan Kim \\
 {\it Department of Physics } \\
{\it Korea Advanced Institute of Science and Technology} \\
{\it 373-1, Kusung-dong, Yusung-ku, Taejon, Korea }
\end{center}
\vspace{2cm}
\begin{abstract}
The effective action is derived for a  self-interacting theory with a
finite fixed  $O(2)$ charge  at finite temperature
in  curved spacetime. We obtain the high temperature expansion  of the
effective action in the weak coupling limit.
In the relativistic temperature,  we discuss about the phase transition in
a homogeneous spacetime.
\end{abstract}

\begin{flushleft}
PACS number(s): 03.70.+k, 04.40.+c, 05.30.Jp
\end{flushleft}

\newpage

\begin{center}
{\bf I.  INTRODUCTION}
\end{center}

Since the discovery of the Bose-Einstein condensation (BEC) much
effort has been made to understand the effect of the non-vanishing
chemical potential on thermodynamical quantities in quantum field
theories in  curved spacetime, where the spatial section may have
a boundary. The high temperature behaviors of the grand thermodynamic
potential in an arbitrary static spacetime have been investigated
by utilizing the heat kernel and zeta-function  regularization
technique \cite{Kirsten}.
 Especially, for an Einstein universe which is a homogeneous space and
have no boundary, the exact results for  the thermodynamical quantities
are derived in the non-relativistic limit and in the relativistic
limit \cite{Singh}.
Recently D. J. Toms studied the BEC in a static spacetime with a possible
spatial boundary and derived the universal character--the portion of
the charge  in the ground state to the  total charge is  identical to
the result in the flat spacetime \cite{Toms}.
But to our
knowledge all of them does not treat the  interaction term.

For the system having the interaction terms, the effective actions  (
potential ) have been only calculated in the flat spacetime.
Kapusta, by studying the effective equation of motion for fields,
showed that charge conservation can affect the phase transition
of a theory \cite{Kapusta}.  Haber and Weldon
studied the effects of a net background charge on ideal and interacting
relativistic Bose gases in the large $N$  approximation \cite{Haber}.
More recently
Benson, Bernstien, and Dodelson exactly calculated  the one-loop finite
temperature effective potential for a self-interacting theory with a
fixed $O(2)$ charge \cite{Benson}.

In this paper we calculate the effective action for a model with an interaction
term and a fixed charge at finite temperature in curved spacetime, and
study the effects of the interaction on the phase transition. The simple model
under investigation,
is described by the matter Lagrangian density
\be
- {\cal L} = \frac{1}{2} ( \partial^\mu \phi_1 \partial_\mu \phi_1
          + \partial^\mu \phi_2 \partial_\mu \phi_2 + m^2 \phi^2 )
	  +  \xi R \phi^2 + \frac{ \lambda}{4 !}  \phi^4,
\ee
where $\xi$ is a numerical factor and $ R$ is the Ricci scalar curvature.
This model is $O(2)$ invariant, and there is  a conserved charge $Q$
associated with $O(2)$ global symmetry.

In Sec. II we set up the formalism for evaluating the effective action
at finite temperature and charge. To evaluate the effective action,
we introduce the Riemann nomal coordinates and use the heat-kernel. In
Sec. III by using the dimensional regularization method, we derive the high
temperature expansion of the effective action in the case that the coupling
constant $\lambda$ is small, and briefly discuss about the renormalization.
In Sec. IV using the result of Sec. III, we discuss about the phase
transition of the theory in the relativistic temperature limit in a
homogeneous spacetime with no boundary. Finally we summary the results.

The spacetime we discuss is the ultrastatic spacetime $ {\cal M} = R \times
\Sigma $.
\be
ds^2 = d \tau^2 + g_{ij} (x) dx^i dx^j,
\ee
where the Wick rotation $\tau = i x^0 $ has to be understood.

\bec
{\bf II.  ONE-LOOP EFFECTIVE ACTION }
\eec

In this section we compute the effective action at finite temperature
and finite charge. According to Ref.\cite{Kapusta,Benson}, the grand partition
function $Z [ \beta, \mu ] = e^{- \beta \Omega ( \beta , \mu ) } $, $\Omega
( \beta , \mu ) $ being the thermodynamic potential, can be expressed as
a path integral:
\be
Z  = N \int_{ \phi_i (\tau = 0,x)  = \phi_i (\tau = \beta,x)}
	   [ d \phi_1 ] [ d \phi_2 ] \exp ( - S [ \phi_i  ] ),
\ee
where $N$ is a constant and
the integration has to be taken over all fields $ \phi ( \tau, x) $ with
periodicity  $\beta = \frac{1}{T}$ with respect to $\tau$.
The action $S$ is
\bea
\nonumber
S &=& \int_o^\beta d \tau \idg \left[ \frac{1}{2} [ \dot{\phi}^2_1 +
\dot{\phi}^2_2 + ( \nabla \phi_1 )^2 + ( \nabla \phi_2 )^2
+  m^2 \phi^2  \right. \\
& & \left.  + \xi R \phi^2 ] + \frac{\lambda}{4 !} \phi^4 +   i
\mu ( \phi_2 \dot{\phi}_1 -
\dot{\phi}_2 \phi_1 ) - \frac{\mu^2}{2} \phi^2 \right] ,
               \label{action}
\eea
where $ \phi^2 = \phi_1^2 + \phi_2^2, \dot{\phi}_i = \frac{ \partial \phi}{
\partial t}$, and $ \mu $ is the Lagrangian multipler related to the
conserved charge $Q$.
It is the same form as the flat spacetime.

 In order to calculate the effective action, some perturbative approach must
 be adopted. The usual one is the loop expansion \cite{Bernard}.
We now expand the action around  a  background field configuration, $\phi_i (x)
 = \bar{\phi}_i (x) $. $ \bph_i $ are determined by
 the equation $ \frac{ \delta \Gamma }{\delta \phi_i } \left|_{ \phi_i
 = \bph_i} = 0  \right.$, where $\Gamma$ is the effective action.
 In general $\bar{\phi}_i (x)$ is not a constant field  in  curved spacetime.
 Then the action in Eq.(\ref{action}) is expanded  in powers of $ \phi_i^{'}
 (x)   = \phi_i (x)  - \bph_i (x) $:
\be
 S = S^{(0)} + S^{(1)} + S^{(2)} + \cdots .
	   \label{ep}
\ee

The zeroth order term is just the action evaluated at $\bar{\phi}_i (x)$:
\be
S^{(0)} = \beta \idg \left[ \frac{1}{2}[  ( \nabla \bph_1 )^2 +
( \nabla \bph_2 )^2 +  m^2 \bph^2 +  \xi R \bph^2 ] + \frac{\lambda}{4 ! }
\bph^4 - \frac{ \mu^2}{2} \bph^2 \right],
\ee
where $\bph^2 = \bph_1^2 + \bph_2^2$. The first order in Eq.(\ref{ep}), which
is linear in $ \phi^{'}_i (x) $, can be neglected. The second order
term in the action is
\be
S^{(2)} = \frac{1}{2} \idgmx \idgmy~ \phi^{'}_i (x) {\cal M}_{i j}
\phi^{'}_j (y),
          \label{ac2}
\ee
where the $2 \times 2$ matrix ${\cal M}_{i j} $ is the second functional
derivative of the action with respect to the field evaluated at $\bph_i$,
\be
{\cal M}_{i j} = \frac{ \delta^2 S}{\delta \phi_i (x)  \delta \phi_j (y) }
    \left|_{\phi_i = \bph_i} \right. .
\ee
Since the boundary condition at finite temperature is periodic for a
bosonic field, we may expand $ \phi^{'}_i (\tau, x ) $ in Fourier modes:
\be
\phi_i^{'} (\tau,x ) = \frac{1}{ \sqrt{\beta}} \sum_{n = -\infty}^{\infty}
\phi^{'}_{i,n} (x) e^{i \omega_n \tau }
\label{fou}
\ee
with  $ \omega_n = \frac{ 2 \pi n }{\beta}$.
Substituting Eq.(\ref{fou}) into Eq.(\ref{ac2}), we obtain
\be
Z^{(1)} = \int [ d \phi^{'}_{1,n}] [ d \phi^{'}_{2,n}] \exp \left[
- \frac{1}{2} \sum_n \idg~ \phi^{'}_{i,n} (x) {\cal M}_{i,j}^{'}
\phi^{'}_{j,n} (x) \right],
\ee
where ${\cal M}_{i,j}^{'}  = ( - \Delta_3 + V_{ij} +\omega^2_n - \mu^2  - 2
\mu \omega_n \epsilon_{ij} )$.
Here $\Delta_3 $ is the Laplacian on the spatial section $\Sigma$.
$V_{ij} $ and $ \epsilon_{ij}$ are given by
\be
V_{ij} =  \xi R \delta_{ij} + \bar{m}^2_1 ( \bph ) \hph_i \hph_j +
\bar{m}^2_2 ( \bph)  ( \delta_{ij} - \hph_i \hph_j ), \\
{}~~\epsilon_{ij} = \left(
		  \begin{array}{cr}
		  0 & -1 \\
		  1 & 0
		  \end{array}
		  \right),
\ee
where
\bea
\bar{m}^2_1 ( \bph ) &=& m^2 + \frac{\lambda}{2} \bph^2, \\
\bar{m}^2_2 ( \bph) &=& m^2 + \frac{ \lambda }{6} \bph^2, \\
\hph_i &=&  \frac{ \phi_i}{ \sqrt{ \phi^2}}.
\eea

In terms of a loop expansion  the effective action is then given by
 \be
       \Gamma [ \bph ] = S^{(0)} [\bph] + \Gamma_{one-loop} + \cdots ,
 \ee
 where the one-loop effective action $ \Gamma_{one-loop}$ is given by
 \be
 \Gamma_{one-loop} = - \ln Z^{(1)}.
 \label{eff}
 \ee
 It is note that the functional $Z^{(1)}$ at finite temperature in
 $d=4$ dimensions is completely reduced to  $3-$dimensional expression.
 Using this fact Eq.(\ref{eff}) can be calculated by the proper time
 formalism in 3-dimensions.

 Following DeWitt \cite{DeWitt}, let us write
 \bea
 \no
 \ln Z^{(1)} &= &\frac{1}{2} \sum_{n = - \infty}^{\infty} \idg \int_0^\infty
 \frac{ds}{s} \langle  x| e^{- s H_n } |x \rangle  \\
 & \equiv& \frac{1}{2} \beta  \idg~ \Gamma
 \eea
 with
 \be
H_n =  - \Delta_3 + V_{ij} + \omega_n^2 - \mu^2 - 2 \mu \omega_n \epsilon_{ij}.
 \ee
 Since the eigenvalues of the operator $H_n$ are not exactly known for a
 general background metric $g$, it is necessary to make use of some
 approximation scheme. The useful approximations are the Riemann normal
 coordinates \cite{Bunch} and the weak field expansion.
  We will use the result of Ref.\cite{Mann}. Following the ansatz for the heat
  kernel suggested by Jack and Parker, we use the nonlocal form
  $ e^{ - s ( \tilde{V}   - \frac{1}{6} R ) } $ in the heat
  kernel \cite{Jack}. The explicit form of  $\Gamma$, up to
  adiabatic order four, is given by ( see Appendix A )
 \bea
 \no
 \Gamma &=& \tr  \frac{1}{\beta} \sum_{n = \infty}^\infty \int_0^\infty
 \frac{ds}{s} \int \frac{d^N p }{(2 \pi )^N } ( I_0 + I_1 +
  I_2 + I_3 + I_4 + \cdots)  \\
 &\equiv& \Gamma_0 + \Gamma_1 + \Gamma_2 + \Gamma_3 + \Gamma_4 + \cdots,
 \eea
 where
 \bea
 I_0 &=& e^{-s (p^2 + \tilde{V} - \frac{1}{6} R )}, \\
 I_1 &=&  0,  \\
 I_2 &=&  - s e^{- s(p^2 + \tilde{V} - \frac{1}{6} R )} \left[
        \frac{1}{6} R - 2 A_{\alpha \beta } ( \delta^{ \alpha \beta } -
        s p^\alpha p^\beta) \right], \\
 I_3 &=&  - i s e^{- s(p^2 + \tilde{V} - \frac{1}{6} R )} \left[
            s  \tilde{V}_{; \alpha} p^\alpha - 3 A_{3 \alpha \beta }
	    ( - s \delta^{\alpha \beta } p^\gamma + \frac{4}{3} s^2
	    p^\alpha p^\beta p^\gamma ) \right], \\
 \no
 I_4  & = &
     -s e^{ - s( p^2 + \tilde{V}- \frac{1}{6} R)}
     \left[
	   \frac{1}{2} \tilde{V}_{:\alpha \beta} (  s \delta^{\alpha \beta}
	  - \frac{4}{3} s^2 p^\alpha p^\beta )
      \right. \\
\no
     & & + \left(   4 A_{ 4 \alpha \beta \gamma \delta} - 2 A_{2 \alpha \beta }
       A_{2 \gamma \delta}
	    \right)
       \left[
	-s ( \delta^{\alpha \beta} \delta^{\gamma \delta} +
	     \delta^{\alpha \gamma} \delta^{\beta \delta} +
	     \delta^{\alpha \delta} \delta^{ \beta \delta} )
	\right. \\
\no
      &&
      - 2s^3 p^\alpha p^\beta p^\gamma p^\delta +
	 \frac{4}{3} s^2 \left( \delta^{\alpha \beta } p^\gamma p^\delta
	  +                \delta^{ \alpha \gamma} p^\beta p^\delta
	              \right.  \\
\no
      && \left. \left. \left.
	  +                \delta^{\alpha \delta} p^\gamma p^ \beta
	  +                 \delta^{\beta \gamma} p^\alpha p^\delta
	  +                \delta^{\beta \delta} p^\gamma p^\alpha
	  +                \delta^{\gamma \delta} p^\alpha p^\beta
	  \right)
	  \right] \right]\\
\no
 & &  +~ s^2 e^{- s(p^2 + \tilde{V} - \frac{1}{6} R )} \left[
	\left[
	\frac{1}{12} R - 2 A_{2 \alpha \beta }
             ( \frac{1}{2}  \delta^{\alpha \beta} - \frac{2}{3}
                   s p^\alpha p^\beta )
        \right]
	\right. \\
\no
 & &  \cdot \left[
             \frac{1}{6} R - 2 A_{2 \alpha \beta} \delta^{\alpha \beta}
       \right]
	     + 2 s A_{2 \alpha \beta } \left[ - \frac{1}{18}
	     R p^\alpha p^\beta   \right. \\
&  &  +  \left.  \left.    \frac{1}{6}
s p^\alpha p^\beta p^\gamma p^\delta R_{\gamma \delta } - \frac{2}{9}
  p^\gamma p^\delta R_{\gamma ~~ \delta}^{~ \alpha ~~~\beta } \right]
  \right].
\eea
 Here $ \tilde{V} = V_{ij} + \omega_n^2 - \mu^2 - 2 \mu \omega_n \epsilon_{ij}
 $.
 After integrating about the momentum $p$,
 we get
 \bea
 \Gamma_0 &=& \tr  \int_0^\infty \frac{ds}{s} \frac{1}{\beta} \sum_n
             \frac{ 1}{ (4 \pi s)^{ \frac{ N}{2} }}
               e^{- s(\tilde{V} - \frac{1}{6} R )  }, \\
\Gamma_1 &=& \Gamma_2 = \Gamma_3 = 0, \\
\no
\Gamma_4 &=& \tr  \frac{1}{\beta} \sum_n  \int_0^\infty
             \frac{ds}{s} \frac{s^2 }{
	     ( 4 \pi s )^{ \frac{N}{2} } }
	     e^{- s(\tilde{V} - \frac{1}{6} R ) } \left[
            -  \frac{1}{6} \Box \tilde{V } \right. \\
      & &   \left.  + \frac{1}{30} \Box R +
	    \frac{1}{180} R_{\alpha \beta \gamma \delta} R^{\alpha \beta
	    \gamma \delta } - \frac{1}{180} R_{\alpha \beta} R^{\alpha \beta}
	    \right],
 \eea
 where $ R, R_{\alpha \beta \gamma \delta}$ and $ R_{\alpha \beta}$ are
 curvatures on the 3-dimensional hypersurface $\Sigma$
 The evaluation of the $\Gamma_0$ and $\Gamma_4$ is not easy because the
 potential $\tilde{V}$ is not a diagonal matrix.
However if we use the property of the trace, we can diagonalized $ e^{- s (
\tilde{V} - \frac{1}{6} R )} $ (see Appendix B). If we choose the matrix $S$ as
 \be
 S = \left(
       \begin{array}{cc}
       ( \bar{m}_1^2 - \bar{m}_2^2 ) \hph_1 \hph_2 - 2 \mu \omega_n &
       ( \bar{m}_1^2 - \bar{m}_2^2 ) \hph_1 \hph_2 - 2 \mu \omega_n  \\
       \lambda_+ - C &      \lambda_- - C
       \end{array}
       \right),
 \ee
 where
 \be
 C =  (\xi - \frac{1}{6}) R + \bar{m}_1^2 \hph^2_1 + \bar{m}_2^2 \hph^2_2
       + \omega_n^2 -\mu^2,
\ee
and $\lambda_{\pm}$ are the eigenvalues of the matrix
$ \tilde{V} - \frac{1}{6} R $:
 \be
 \lambda_{\pm} = (\xi - \frac{1}{6}) R + m^2 + \frac{\lambda}{3} \bph^2 +
 \omega_n^2 - \mu^2 \pm \frac{1}{2} \left[ \frac{ \lambda^2 }{9}
 \bph^4 - 16 \mu^2 \omega^2_n \right]^{\half}.
 \ee
Then we can rewritten the $\Gamma_0$ and $\Gamma_4$ as a more tractable form:
 \bea
 \Gamma_0 &=&  \frac{1}{\beta} \sum_n ( 4 \pi )^{-N/2} \int_0^\infty
 \frac{ds}{s} s^{-N/2 } \tr
		  \left(
		  \begin{array}{cc}
		    e^{- s \lambda_+ }   &  0   \\
		    0                    &  e^{-s \lambda_- }
		  \end{array}
		  \right),
		  \\
\no
 \Gamma_4 &=&  \frac{1}{\beta} \sum_n ( 4 \pi )^{-N/2} \int_0^\infty
 \frac{ds}{s} s^{-N/2 +2 } \tr
		  \left(
		  \begin{array}{cc}
		    e^{- s \lambda_+ }   &  0   \\
		    0                    &  e^{-s \lambda_- }
		  \end{array}
		  \right)
		  \\
\no
 &&  \cdot\left[ - \frac{1}{6} S^{-1} \Box \tilde{V} S + \tilde{a}_2 \right]\\
\no
 &=&  \frac{1}{\beta} \sum_n ( 4 \pi )^{-N/2} \int_0^\infty \frac{ds}{s}
  s^{-N/2 +2 }
  \left\{
	( e^{-s \lambda_+ } + e^{- \lambda_-}) ( \tilde{a}_2
	- \frac{\lambda}{18} \Box \bph^2  )
	\right. \\
\no
& &  - \frac{\lambda}{18}
	\left[
	     \frac{\lambda^2}{9} \bph^4 - 16 \mu^2 \omega_n^2
        \right]^{-\half} \left[ \frac{\lambda}{6} ( \bph^2_1 - \bph^2_2 ) \Box
(
	    \bph_1^2 - \bph_2^2 )  \right.\\
 &+&  \left. \left.\frac{2 }{3} \lambda \bph_1 \bph_2
	    \Box ( \bph_1 \bph_2 ) \right]
	    ( e^{- s \lambda_+} - e^{-s \lambda_- } )
	    \right\},
 \eea
 where $ \tilde{a}_2 = \frac{1}{30} \Box R +
	    \frac{1}{180} R_{\alpha \beta \gamma \delta} R^{\alpha \beta
	    \gamma \delta } - \frac{1}{180} R_{\alpha \beta} R^{\alpha \beta}
    $.

 Now, let us consider limiting case.
 First of all, we consider the case that the metric is flat.
 In flat spacetime  the background fields $\bph_i$ is a constant because
 the spacetime is homogeneous.
Then only $\Gamma_0$ survive. Using the formula
\bea
\no
 G_0 &=& \frac{ \partial \Gamma_0 }{\partial m^2} \\
\no
     &=& -  \frac{1}{\beta} \sum_n \int \frac{ d^N p}{ (2 \pi)^N}
     \int_0^\infty ds ( e^{- s ( p^2 +\lambda_+)} + e^{- s( p^2 + \lambda_-)} )
\\
\no
  &=& -  \frac{1}{\beta} \sum_n \int \frac{ d^N p}{ (2 \pi)^N}
   \left( \frac{1}{ p^2 +\lambda_+} + \frac{1}{ p^2 +\lambda_-} \right) \\
\no
  &=& -  \frac{1}{\beta} \sum_n \int \frac{ d^N p}{ (2 \pi)^N}
  \frac{\partial}{\partial m^2} \ln \left[ [p^2 +( \xi - \frac{1}{6})R + m^2 +
  \frac{\lambda}{3} \bph^2  \right.\\
  & &  \left. + \omega_n^2 - \mu^2 ]^2 - \frac{1}{4} [ \frac{ \lambda^2}{9}
  \bph^4 - 16 \mu^2 \omega_n^2 ] \right],
\eea
we obtain, after sum over $n$,
\bea
\no
\Gamma_0 &=& -  \frac{1}{\beta} \int \frac{d^N p}{(2 \pi )^N} \left[
	 \beta E_+  +  2 \ln ( 1 + e^{- \beta E_+ } )  \right. \\
     & &  \left. + \beta E_- + 2 \ln (
	 1 + e^{- \beta E_-} ) \right],
\eea
where $ E_{\pm}^2 = p^2 + M_f^2 + \mu^2 \pm [ 4 \mu^2 ( p^2 + M_f^2 )^{1/2}
+ \frac{ \lambda^2}{36} \bph^4 ]^{1/2}$ with
$M_f^2 = m^2 ( 1+ \frac{\lambda}{ 3 m^2} \bph^2 )$.
In this case, the effective action ( potential) is given by
\bea
\no
\Gamma[ \bph] &=& S^{(0)} [ \bph] + \beta \idg \frac{1}{\beta} \int
\frac{ d^N p}{( 2 \pi)^N} \left[
	 \frac{\beta( E_+ + E_- )}{2} \right. \\
 & &     + \left. \ln ( 1+ e^{-\beta E_+} ) +\ln ( 1+ e^{- \beta E_-} )
 \right].
 \eea
This is exactly the same form as the previous result in the flat
spacetime \cite{Benson}.

Second, for the case $\lambda \rightarrow 0, \lambda_\pm \rightarrow
( \xi -\frac{1}{6}) R + m^2 + ( \omega_n \pm i \mu )^2$.
Then $\Gamma_0$ and $\Gamma_4$ go to the previous result because $ \Box \V
\rightarrow 0$.
For $ \mu = 0$, $\lambda_+ = (\xi - \frac{1}{6}) R + m^2 +
\omega_n^2 + \frac{1}{2}
\lambda \bph^2 $ and $ \lambda_- = (\xi - \frac{1}{6}) R + m^2 +
\omega^2_n + \frac{1}{6}
\lambda \bph^2$. In this case  the effective action goes to the previously
known result \cite{Kirsten}.

\bec
{\bf III. HIGH-TEMPERATURE EXPANSION }
\eec

In this section we will calculate $ \Gamma_0$ and $ \Gamma_4 $ in the high
temperature limit and the coupling constant $\lambda$ is small.
The regularization method is the dimensional one \cite{Hu}. After the
regularization, we take $ N \rightarrow 3 $.
Fisrt of all, let us calculate  $ \Gamma_0 $. For small $
\lambda$, $ \Gamma_0$ can be written as
\bea
\no
& \Gamma_0 & \\
\no
&=&  \frac{1}{\beta} \sum_{n=-\infty}^\infty (4 \pi )^{- N/2} \Gamma
(- \frac{N}{2} ) \left\{
		 \left[  M^2 + \omega_n^2 - \mu^2 + \frac{1}{2} \left(
	 \frac{ \lambda^2}{9} \bph^4 - 16 \mu^2 \omega_n^2 \right)^{
	 \frac{1}{2}}
		 \right]^{\frac{N}{2}}
		 \right. \\
\no
	 & & + \left. \left[
	M^2 + \omega_n^2 - \mu^2 - \frac{1}{2} \left( \frac{\lambda^2}{ 9}
		\bph^4 - 16 \mu^2 \omega^2_n \right)^{\half}
		\right]^{\Nhalf}
	       	\right\} \\
&=& \Gamma_0^{(n=0)} + \Gamma_0^{(1)}  + \Gamma_0^{(2)} + \cdots,
\eea
where
\bea
\no
\Gamma_0^{( n=0)} &=&  \frac{1}{\beta} (4 \pi )^{-N/2} \Gamma ( - \frac{N}{2} )
        \left\{
	   \left[ M^2 - \mu^2 + \frac{\lambda}{6} \bph^2 \right]^{\Nhalf}
	   \right.  \\
&& \left.   +\left[ M^2 - \mu^2 - \frac{\lambda}{6} \bph^2 \right]^{\Nhalf}
        \right\},
	\label{gaa} \\
\no
\Gamma_0^{(1)} &=&  \frac{1}{\beta} (4 \pi )^{-N/2} \Gamma ( - \frac{N}{2} )
	   2 \sum_{n=1}^\infty
        \left\{ \left[ M^2 + ( \omega_n - i \mu )^2 \right]^{\Nhalf}
	   \right. \\
   &&  \left.
              +  \left[ M^2 + ( \omega_n + i \mu )^2 \right]^{\Nhalf}
	\right\},
	\label{gab} \\
\no
\Gamma_0^{(2)} &=&  \frac{1}{\beta} (4 \pi )^{-N/2} \Gamma ( - \frac{N}{2} )
	   2 \sum_{n=1}^\infty
        \left\{   i \frac{N}{2} \frac{ \lambda^2 \bph^4}{ 144 \mu \omega_n}
		\left[ M^2 + ( \omega_n - i \mu )^2 \right]^{\Nhalf -1 }
		\right. \\
		&& \left.
          - i \frac{N}{2} \frac{ \lambda^2 \bph^4}{ 144 \mu \omega_n}
		\left[ M^2 + ( \omega_n + i \mu )^2 \right]^{\Nhalf -1 }
	\right\},
	\label{gac}
\eea
where $M^2 = m^2 - \mu^2 + \xi R + \frac{\lambda}{6} \bph^2 $.
The evaluations of the Matsubara sums in Eqs.(\ref{gaa},\ref{gab},\ref{gac})
are not difficult \cite{Actor}.
By using the dimensional regularization, we obtain
\bea
\Gamma_0^{(0)} &=& 	 \frac{1}{\beta} \frac{1}{6 \pi} \left[
	  ( M^2 - \mu^2 + \frac{\lambda}{6} \bph^2 )^{3/2}
	  + ( M^2 - \mu^2 - \frac{\lambda}{6} \bph^2 )^{3/2}
		      \right],  \\
\no
\Gamma_0^{(1)} &=& 	 2 \left[
	  \frac{\pi^2 }{45} \frac{1}{\beta^4} - \frac{1}{12 \beta^2} (
	  M^2 - 2 \mu^2)  +
	  \left[ \pole  \right] \frac{ M^4}{( 4 \pi)^2 }
	  \right.  \\
\no
     &&  + \left.  \frac{ \mu^2}{8 \pi^2} ( M^2 - \frac{1}{3} \mu^2 ) +
	 \frac{ \pi^2}{ \beta^4} \sum_{k,r \geq 1 , k+r \neq 2 }^\infty
	 \frac{8}{3}  C( - \frac{3}{2}, k,2r ) (-1)^r
	  \right. \\
      &&  \left.
	  ( \frac{\mu \beta}{2 \pi} )^{ 2r} ( \frac{ M \beta}{2 \pi} )^{2k}
	  \zeta ( -3 + 2 k + 2 r )
	 \right],  \\
\no
\Gamma_0^{(2)} &=&   \frac{ \lambda^2 \bph^4 }{ 288 \pi^2 } \left[
	 \left[ \pole \right]
	 + \sum_{k,r \geq 0, k+r \neq 0 }^\infty C( -\frac{1}{2}, k,2r+1)
         \right.	 \\
	 && \left.
 (-1)^r (\frac{ \mu \beta }{2 \pi} )^{2r} ( \frac{ \beta M}{2 \pi } )^{2k}
	 \zeta ( 2k + 2r + 1 )
		   \right],
\eea
where $ \epsilon = 4 - d $ and $\gamma$ is the Euler constant $0.577...$,  and
\be
C( s - \frac{N}{2} , k,r) = (-1)^{k+r} \frac{
	\Gamma( s - \frac{N}{2} + k ) \Gamma ( 2s -N +2k +2r ) }{
	k! r! \Gamma( s - \frac{N}{2} ) \Gamma( 2s - N + 2k) }.
\ee

Now let us calculate the $\Gamma_4$. After some algebra, we get
\bea
\no
\Gamma_4 &=& \frac{1}{4 \pi^2} \left\{
                  	\left[ \pole \right]
  + \sum_{ k,r \geq 0, k+r \neq 0} C ( \frac{1}{2},k,2r)  \right. \\
\no
& &  \cdot  \left. ( \frac{ i \mu \beta}{2 \pi } )^{2r}
	  ( \frac{ \beta M}{2 \pi}^{2k} \zeta ( 1 + 2k + 2r)
      \right\}
       ( \tilde{a}_2 - \frac{\lambda}{18} \Box \bph^2 ) \\
\no
& + \frac{1}{8 \pi \beta} &
   \left\{
      \left[
	 \left[ \Mp \right]^{-\half}
	 + \left[ \Mm \right]^{-\half}
      \right]
       ( \tilde{a}_2 - \frac{\lambda}{18} \Box \bph^2 )
   \right.    \\
\no
& & -  \frac{\lambda}{6 \bph^2} \left( \Kit  \right) \\
\no
&&  \cdot \left. \left[ \left[  \Mp \right]^{-\half} -
		 \left[  \Mm \right]^{-\half}   \right]
   \right\}	 \\
\no
& & + \frac{\lambda^2}{288 \pi^2 }
  \left[   \Kit \right]  \\
&& \cdot \sum_{k,r \ge 0}  C ( \frac{1}{2} , k,2r +1) (-1)^{2r}
   ( \frac{  \mu \beta}{ 2 \pi })^{2r}  ( \frac{\beta M}{2 \pi} )^{2k}
	      \zeta ( 3 + 2k + 2r ),
\eea
 where we have calculated up to $ \lambda^2 $ order except  $ n=0$
 term.
 The final result for the effective action is given by
 \be
 \Gamma[ \bph] = S[\bph] - \frac{1}{2} \beta \idg
 ~( \Gamma_0^{(n=0)} + \Gamma_0^{(1)} + \Gamma_0^{(2)} + \Gamma_4  + \cdots)
 \label{ef2}
 \ee
 up to $\lambda^2$ order.

 Let us consider the infinite part of the effective action.
 It is given by
 \bea
 \no
 \Gamma_{infinite} [ \bph] &=& - \idg 2 \frac{1}{ (4 \pi)^2}
	      \left( \polea \right)  \\
 && \left[  \frac{M^2}{2} + \frac{\lambda^2 }{72} \bph^4 + (
    \tilde{a}_2 - \frac{\lambda}{18} \Box \bph^2 ) \right],
\eea
where $\tilde{\mu}$ is a arbitrary mass scale.
It is note that the value of the pole part differs from one that calculated
by the zeta function regularization method \cite{Kirsten,Actor}.
In the case $ \lambda \rightarrow 0 $, this reduce to the known result.
These infinite terms can be removed by absorbing it into the
gravitational Lagrangian and by introducing the counter terms \cite{Birrel}.
Up to terms which are total divergence, the coupling constant and mass
counter terms are only needed. This is coincide with the fact that
at one loop level, there is no wave functional counter term \cite{Toms2}.

\bec
{\bf IV. BOSE-EINSTEIN CONDENSATION}
\eec

In quantum field theory, BEC is interpreted as a symmetry breaking effect
in flat as well as in curved spacetime \cite{Toms,Haber,Benson}.
In order to discuss the BEC in the case of a relativistic Bose gas,
it is important to clarify in what sense we mean that the temperature is
relativistic. In flat spacetime, it means that $T \gg M $. In curved
spacetime, in addition to $T \gg M $ , we  require $ T \gg |R|^{1/2}, $
where $|R|$ is the magnitude of a typical curvature of the
spacetime \cite{Toms}.

 The high temperature effective action, from Eq.(\ref{ef2}),
 has the form in the relativistic temperature,
 \bea
 \no
 \Gamma [ \bph] &=& S[ \bph] + \Gamma_{one-loop} \\
 \no
 &=&     \beta \idg \left[
	      \frac{ 1}{2} \nabla \bph \cdot \nabla \bph + \frac{1}{2}
	      m^2 \bph^2 + \frac{1}{2} \xi R \bph^2 + \frac{\lambda}{4!}
	       - \frac{\mu^2}{2} \bph^2  \right. \\
	&& \left. - \frac{\pi^2}{45} \frac{1}{\beta^4} + \frac{1}{12 \beta^2}
	   ( M^2 - 2 \mu^2 ) + \cdots \right].
\eea
At first we discuss about the case that the spacetime is homogeneous. The
 translational invariance  implies that the background field is a constant
( This is always not true, For $\lambda = 0$, see Ref. \cite{Toms,Gognola}).
 In this case, the equation of motion for $\bph_i$ implies
 \be
 ( m^2 - \mu^2 + \xi  R + \frac{\lambda}{6} \bph^2 + \frac{\lambda}{18}
 T^2 ) \bph_i =0 .
 \ee
Thus the effective action have two minima: $\bph_i =0 $, with unbroken
symmetry,
 \bea
 \no
 \bph^2 &=& \frac{6}{\lambda} \left[
		\mu^2 - (m^2 + \xi R  + \frac{\lambda}{18} T^2) \right] \\
   & =& \frac{6}{\lambda} \left[ \mu^2 - m^2 (T) \right]
   \label{bro}
\eea
with broken symmetry.
 This is the same form as the result in the flat spacetime except for
 changing $ m^2 \rightarrow
 m^2 + \xi R $ \cite{Benson}.
 The critical temperature when  phase transition occurs is
 \be
 T_c^2 = \frac{18}{\lambda} \left[ \mu^2(T) -( m^2  + \xi R ) \right]
 \ee
 In general, $\mu^2$ has the temperature dependence due to the
 charge conservation.

 Now consider the expectation value of the charge operator $Q$ which
 is given in terms of the effective action by
  \bea
  \no
  Q &=& - \frac{1}{\beta} \frac{ \partial \Gamma}{ \partial \mu} \equiv
  Q_0 + Q_{th} \\
  &=& \idg ( \mu \bph^2 + \frac{\mu}{3} T^2 )
  \label{charge}
  \eea
  in the high temperature approximation.
  $Q_0$ is the charge in the ground state and $Q_{th}$ is the charge in the
  thermal excited state.
   If $T$ is high, $ \bph_i =0 $. Then $Q = \frac{\mu }{3} T^2 V$, where
   $V$ is the volume of the spatial section $\Sigma$.
 As   $T$ decreases, $\mu$ must increase until the temperature becomes $T_c$.
 From Eq.(\ref{bro}) and Eq.(\ref{charge}) the critical temperature is
 rewritten as
 \be
 T_c = \left( \frac{3 Q}{V} \right)^{1/2} ( m^2 (T))^{-1/2}.
 \ee
 For $T \leq T_c$, it is easily seen that
\be
	Q_0 = Q \left[ 1 - ( \frac{T}{T_c} )^2 \right],
	\label{qcha}
\ee
which is identical form to the result in flat spacetime \cite{Benson}.

Up to date we
have restricted the value of the fields $\bph_i$ to be a constant.
But in curved spacetime or in the case that there are boundaries,
as others showed,
the value of the  fields $\bph_i$ is not a costant even for $\lambda = 0$.
In addition, the  critical temperature differs from  our result because
there is a non-zero lowest eigenvalue for the Laplacian operator
\cite{Toms,Gognola}.

 In an inhomogeneous spacetime, the equation of motion for the fields $
 \bph_i$ is given by
 \be
 \Box \bph_i - ( m^2 - \mu^2 + \xi R + \frac{1}{6} \bph^2 +
 \frac{\lambda}{18} T^2 ) \bph_i = 0.
 \ee
This is a nonlinear equation, so it is not easy to find the analytic solution.
Therefore it is difficult to discuss about the phase transition of the system.

\bec
{\bf V.   DISCUSSION}
\eec
 In this paper we have calculated the effective action for the $O(2)$
 invariant model with quartic self-interaction and global $O(2)$ fixed charge
 at finite temperature in  curved spacetime. We calculated the effective
 action by using the modified heat
 kernel technique suggested by Jack and Parker and introducing
 the normal coordinates expansion method.
 We also used the fact that a $2 \times 2 $ matrix can be diagonalized
 in general.  We showed that
 the coefficient $a_1$  - the Minakshiuundaram-Seeley- DeWitt coefficient-
 does not appear similarly to the result at  zero temperature case.

Using the effective action, we discussed about the phase transition in the
homogeneous spacetime.
 It is shown that the charge condensation property is  the same as the
 flat spacetime case, except for the changing of the mass. The form of
 the condensated
 charge fraction to the total charge is the same as the non-interacting
 case (Eq.(\ref{qcha}).
 For the  inhomogeneous spacetime we have not discussed about the phase
 transition because the equation of motion is nonlinear.

  Even if we have studied  the simple $O(N=2)$ model,  our result can
  be generalized to an arbitrary potential type and to the $N >2$ model.
  But for $N>2$ one will be confronted with the diagonalization problem
  of the matrix.

For conformally ultrastatic spacetime the effective action may also be
derived by using the conformal transformation technique.

\bec
{\bf ACKNOLEDGEMENT}
\eec
This work was  partially supported by KOSEF ( Korea Science and Engineering
Foundation)

\vspace{0.5cm}

\appendix
\bec
{\bf APPENDIX  A}
\eec
In normal coordinates, an operator ${\cal O } (x,x') $ may be expanded in
powers of $y$ about the point $x'$, where $ \left. y^\mu =
\frac{d x^\mu (\tau)}{d \tau}
\right|_{\tau = 0}, x( \tau) $ being a geodesic connecting $x$
and $x'$ \cite{Bunch}.
Then, up to adiabatic order four \cite{Mann},
\be
( - \Box + \V ) = C_1 + C_2 + C_3 +C_4 + \cdots,
\ee
where
\bea
C_1 &=& ( - \partial^2 + \V - \frac{1}{6} R ),\\
C_2 & =&  - ( A_{2 \alpha \beta} y^\alpha y^\beta + B_{2 ~ \alpha \beta}^{
\mu \nu} \partial_\mu y^\alpha y^\beta \partial_\nu - \frac{1}{6} R ),\\
C_3 &=&  - ( A_{3 \alpha \beta \gamma} y^\alpha y^\beta y^\gamma \partial^2
	+ B_{2 ~ \alpha \beta \gamma}^{\mu \nu} \partial_\mu y^\alpha
	y^\beta y^\gamma \partial_\nu - \V_{:\alpha} y^\alpha ),\\
\no
C_4 &=& - ( A_{2 \alpha \beta} B_{2~ \gamma \delta }^{\mu \nu} y^\alpha
y^\beta \partial_\mu y^\gamma y^\delta \partial_\nu + A_{4 \alpha
	  \beta \gamma \delta} y^\alpha y^\beta y^\gamma y^\delta \partial^2
	  \\
&&       + ~B_{4 ~\alpha \beta \gamma \delta }^{\mu \nu} \partial_\mu y^\alpha
 y^\beta y^\delta y^\gamma \partial_\nu - \frac{1}{2} \V_{:\alpha \beta}
	  y^\alpha y^\beta ),
\eea
with
\bea
A_{2 \alpha \beta} &=& \frac{1}{6} R_{\alpha \beta}, \\
A_{3 \alpha \beta \gamma} &=& \frac{1}{12} R_{\alpha \beta :\gamma}, \\
A_{4 \alpha \beta \gamma \delta} &=& \frac{1}{40} R_{\alpha \beta :\gamma
\delta} + \frac{1}{180} R_{\mu \alpha \beta }^{~~~~\nu}
R_{~\gamma \delta \nu}^{\mu}
 + \frac{1}{72} R_{\alpha \beta }R_{\gamma \delta}, \\
B_{2~ \alpha \beta}^{\mu \nu} &=& \frac{ 1}{3} R_{~\alpha~ \beta}^{\mu~\nu}
 - \frac{1}{6} \delta^{\mu \nu} R_{\alpha \beta}, \\
B_{3 ~ \alpha \beta \gamma}^{ \mu \nu}   &=& \frac{1}{6}
R_{~\alpha ~ \beta :\gamma}^{\mu~ \nu} -
\frac{1}{12} \delta^{\mu \nu} R_{\alpha \beta: \gamma }, \\
\no
B_{4 ~ \alpha \beta \gamma \delta }^{\mu \nu} &=&
	\delta^{\mu \nu} \left(
    - \frac{1}{40} R_{\alpha \beta:\gamma \delta} - \frac{1}{180}
    R_{\sigma \alpha \beta }^{~~~~\lambda} R_{~\delta \gamma \lambda}^{
    \sigma} + \frac{1}{72} R_{\alpha \beta} R_{\gamma \delta} \right) \\
     & & + \left(
	    \frac{1}{20} R_{~\alpha ~ \beta :\gamma \delta }^{\mu~\nu}
	    + \frac{1}{15} R_{~\alpha \lambda \beta }^{\mu}
	    R_{~\gamma ~ \delta}^{ \lambda ~ \nu} -
	    \frac{1}{18}  R_{\alpha \beta} R_{~\gamma ~ \delta}^{\mu ~ \nu}
	    \right)
\eea
By using the Schwinger expansions \cite{Sch}
\bea
e^{-( H_0+ H_I)s } &=& e^{-H_0 s} + (-s) \int_0^1 du e^{-(1-u)H_0 s }
	H_I e^{-u H_0 s }  \\
\no
&+& (-s)^2   \int_0^1 du u \int_0^1 dv e^{-(1-u)H_0 s } H_I e^{ -
	u(1-v) H_0 s } H_I e^{-uv H_0 s} + \cdots ,
\eea
one may evaluate the heat kernel
\bea
\no
K &=&  e^{ (- \Box + \V)s}   \\
\no
  &=& e^{ - ( - \partial^2 + \V - \frac{1}{6} R ) s }
 + (-s) \int_0^1 du   e^{ - (1-u)( - \partial^2 + \V - \frac{1}{6} R ) s }
 \\
\no
 && \cdot ( C_2 + C_3 + C_4 ) e^{ - u( - \partial^2 + \V - \frac{1}{6} R ) s }
 \\
\no
 && +~ (-s)^2 \int_0^1 du u \int_0^1 dv e^{ -(1-u) ( - \partial^2 + \V -
 \frac{1}{6} R ) s } C_2
  e^{ -u(1-v) ( - \partial^2 + \V - \frac{1}{6} R ) s }
  \\
 && \cdot C_2 e^{ - uv( - \partial^2 + \V - \frac{1}{6} R ) s }
   + \cdots .
\eea
In local momentum space, for any operator $ {\cal O } (x,x')$
\bea
\no
\langle  x | {\cal O} ( x,x') | x' \rangle  &=& \langle y |{\cal O} ( y, x') |
	   0 \rangle \\
&=& \int dp dq \langle y| p \rangle \langle p|  {\cal O} ( i
     \frac{\partial}{\partial p} , x' ) | q \rangle \langle q| x \rangle,
\eea
where
$y = x-x'$.
Then, using the trace property $\tr (ABC) = \tr (CAB) $, one can obtain
\be
\tr \langle x | K | x \rangle
= \tr \int \frac{d^N p}{(2 \pi )^N } ( I_0 + I_1 + I_2 + I_3 + I_4 + ...) .
\ee


\bec
{\bf APPENDIX B}
\eec
For any real $2 \times 2$ matrix A, a  matrix $S$ can be chosen so that
$ S^{-1} A S $ has the diagonal form.
Let
\be
 A = \left( \begin{array}{cc}
	  a & b \\
	  c & d
	  \end{array}
	  \right), ~~
 A  { \alpha  \choose \gamma } = \lambda_+ { \alpha \choose \gamma  },
  ~~ A {\beta \choose \delta } = \lambda_- {\beta \choose \delta},
\ee
where $\lambda_\pm $ are eigenvalues, and $ {\alpha \choose \gamma } $ and
${\beta \choose \delta } $ are the corresponding eigenvectors.
They  are given by
\bea
\lambda_\pm &=& \frac{1}{2} \left[
       ( a+ d) \pm \left( ( a+d )^2 - 4 ( ad - bc) \right)^{1/2} \right]
	      , \\
	       {\alpha \choose \gamma } & = & {b \choose \lambda_+ - a },
	       ~~ {\beta  \choose \delta } + { b \choose  \lambda_- - a }.
\eea
If we choose the matrix $S$ as
\be
    S = \left(    \begin{array}{cc}
	    \alpha & \beta \\
	    \gamma & \delta
	    \end{array}
	    \right),
\ee
then
\be
S^{-1} A S = \left(
	     \begin{array}{cc}
	     \lambda_+  & 0 \\
	     0         &   \lambda_-
	     \end{array}
	     \right).
\ee

 From this we can show that
\bea
S^{-1} e^{t A} S &=& e^{ S^{-1} A S } = \exp
	    \left(   \begin{array}{cc}
		\lambda_+ t & 0 \\
		0           & \lambda_-
		\end{array}
            \right)
	     =
	    \left(   \begin{array}{cc}
	     e^{ \lambda_+ t}  & 0 \\
	     0                 &  e^{ \lambda_- t }
		\end{array}
            \right), \\
 \tr e^{At} &=& \tr ( S^{-1} e^{At} S )  = ( e^{\lambda_+ t } +
 e^{\lambda_- t }),  \\
\no
 \tr e^{At} B &=& \tr ( S^{-1} e^{At} S S^{-1} B S ) \\
	      &=&  \tr \left[
			 \left( \begin{array}{cc}
			 e^{\lambda_+ t } & 0 \\
			 0                &  e^{\lambda_- t }
			 \end{array}
			 \right)
			 S^{-1} B S
			 \right].
\eea

\vspace{0.5cm}


\end{document}